\documentclass{PoS}
\usepackage{amsmath,amssymb}
\usepackage{times}
\usepackage{delarray}
\usepackage{rotating}

\title{Extrapolation Algorithms for Infrared Divergent Integrals}

\ShortTitle{Extrapolation for IR Divergent Integrals}
\vspace*{-0.8cm}

\author{\speaker{Elise de Doncker} \\
        Western Michigan University\\
        E-mail: \email{elise.dedoncker@wmich.edu}}
\author{Junpei Fujimoto\\
        High Energy Accelerator
Research Organization (KEK)\\
        E-mail: \email{junpei.fujimoto@kek.jp}
}
\author{Nobuyuki Hamaguchi\\
        Hitachi ICT Business Services, Ltd., Japan\\
        E-mail: \email{nobuyuki.hamaguchi.xr@hitachi.com}
}
\author{Tadashi Ishikawa\\
        High Energy Accelerator
Research Organization (KEK)\\
        E-mail: \email{tadashi.ishikawa@kek.jp}
}
\author{Yoshimasa Kurihara\\
        High Energy Accelerator
Research Organization (KEK)\\
        E-mail: \email{yoshimasa.kurihara@kek.jp}
}
\author{Marko Ljucovic\\
        Western Michigan University\\
        E-mail: \email{markoljucovic@gmail.com}
}
\author{Yoshimitsu Shimizu\\
        High Energy Accelerator Research Organization (KEK) \\
        E-mail: \email{yohimitsu.shimizu@kek.jp}
}
\author{Fukuko Yuasa\\
        High Energy Accelerator
Research Organization (KEK)\\
        E-mail: \email{fukuko.yuasa@kek.jp}
}
\abstract{This paper describes applications of extrapolation for the computation
of coefficients in an expansion of infrared divergent integrals. 
An extrapolation procedure is performed with respect to a parameter introduced by 
dimensional regularization. While this treats typical IR singularities at the 
boundaries of the integration domain, special care needs to be taken in cases where
the integrand is singular in the interior of the domain as well as on the boundaries.
A double extrapolation is devised for a class of massless vertex integrals.
Quadruple precision results are presented, demonstrating high accuracy.
The computations are supported by the use of general adaptive integration programs
from the {\sc Quadpack} package, in iterated integrations with highly singular integrand
functions.
}

\FullConference{3rd Computational Particle Physics Workshop \\
                 September 23-25, 2010\\
                 Tsukuba, Japan}

\begin{document}

\section{Introduction}
\label{intro}
\noindent
For \emph{infrared (IR)} divergent loop integrals, the integrand functions have non-integrable
singularities through vanishing denominators. Based on dimensional regularization, the
integral is expanded as a function of a parameter that approaches zero~\cite{kurihara05a}.

In this paper we report numerical results for the leading coefficients obtained
by convergence acceleration (extrapolation) of a sequence of integral approximations.
The methods are explained in detail in~\cite{jocs11}, where also results are given using
double precision arithmetic. In the present paper we give extensive results using
quadruple precision and show that in many cases the accuracy can be improved to near the
relative machine accuracy. 
 
According to the asymptotic behavior of the integral, we can explore
linear or nonlinear extrapolation techniques.
The asymptotic expansion gives rise to linear systems of the form
\begin{equation}
S_\ell = \sum_{k=0}^{n-1} a_k \,\varphi_k(\varepsilon_\ell), ~~1\le\ell\le n,
\label{linsys}
\end{equation}
where $S_\ell$ is generally a scaled version of the integral $I(\varepsilon),$ approximated numerically.
A linear system solver or a linear extrapolation method can be used if the $\varphi_k$ 
are known functions of $\varepsilon_\ell.$ Otherwise, a nonlinear extrapolation may be suitable,
depending on the nature of the $\varphi_k$ functions.

For (small) $\varepsilon > 0,$ the numerical integral approximation
may be affected by singular integrand behavior which occurs at the boundaries and/or 
in the interior of the integration domain.
The sample integrals in this paper pertain to classes of one-loop vertex integrals which are two-dimensional,
over the unit triangle $\{\, (x,y) ~|~ 0 \le y \le 1-x \le 1 \,\}.$
An iterated or repeated numerical integration can be performed efficiently with the 
general adaptive integration programs {\sc Dqags/Dqag} from {\bf{\sc Quadpack}}~\cite{pi83}.

The expansions derived symbolically in~\cite{kurihara05a} generally involve hypergeometric
functions. 
We calculate the hypergeometric function numerically in Section~\ref{hyper}, using an extrapolation 
to handle the singularity in the integration interval.
In Section~\ref{one-off-shell} we present results for the case of 
one off-shell ($p_3^2 \ne 0$) and two on-shell ($p_1^2 = p_2^2 = 0$) particles. 
The coefficients of the divergent terms in the integral expansion are calculated with
an extrapolation as the parameter $\varepsilon$ introduced by dimensional regularization 
goes to zero. The integrals in the sequence have integrand singularities on the boundaries
of the integration region.

Section~\ref{two-off-shell} addresses IR divergent integrals with one on-shell and two
off-shell particles, where integrand singularities may occur in the interior as
well as on the boundaries of the integration domain. In this case, the integrals in the
extrapolation sequence with respect to $\varepsilon$ involves an extrapolation to deal with
the interior singularity.


\section{Extrapolation}
\label{secext}
\noindent
For a sequence $\{S(\varepsilon_\ell)\},$ which converges 
to the limit $\mathcal S = \lim_{\varepsilon_\ell \rightarrow 0} S(\varepsilon_\ell),$
an extrapolation may be performed with the goal of creating sequences which convergence
faster than the given sequence, based on its asymptotic expansion
\begin{equation}
S(\varepsilon) \sim {\mathcal S} + A_1\varphi_1(\varepsilon) +
A_2\varphi_2(\varepsilon) + \ldots.
\label{asym}
\end{equation}
as $\varepsilon\rightarrow 0.$
In the context of series convergence we consider the limit of its partial sums. Some
extrapolation methods allow summing divergent series to a value referred to as 
anti-limit.

A linear extrapolation yields solutions to linear systems of the form 
\begin{equation}
S(\varepsilon_\ell) = a_0 + a_1\varphi_1(\varepsilon_\ell) +
 \ldots a_\nu\varphi_\nu(\varepsilon_\ell), ~~~\ell = 0,\ldots,\nu,
\label{exp}
\end{equation}
of order $(\nu + 1)~ \times ~(\nu + 1)$ 
for increasing values of $\nu$~\cite{lyness76,brezinski80}.
The sequence of $\varepsilon_\ell$ may be geometric or another type of sequence that
decreases to 0. As an example, Romberg integration relies on the Euler-Maclaurin 
expansion of the integral as a function of the step size $\varepsilon = h.$ Then
(\ref{asym}) is assumed to be an expansion in even powers of $h,$ for the composite 
trapezoidal rule values $S(h)$ with $h = 2^{-\ell}, ~\ell \ge 0.$ Values for
$a_0 \approx {\mathcal S}$ are obtained for successive $\nu$ by solving the 
$(\nu + 1)~ \times ~(\nu + 1)$ systems of~(\ref{exp}) implicitly using the Neville
algorithm. 

More general sequences of $\varepsilon$ include the sequence by Bulirsch, of
the form $1/b_\ell$ with $b_\ell = 2, 3, 4, 6, 8, 12, \ldots .$ (consisting of powers
of 2, alternating with $1.5\times$ the preceding power of 2). The type of sequence selected 
 influences
the stability of the process, which was found more stable with the geometric sequence
than with the harmonic sequence (with the Bulirsch seuence in between)~\cite{lyness76}.
On the other hand there is a trade-off with the computational expense of $S(\varepsilon),$
which may become prohibitive for fast decreasing $\varepsilon.$ For the computations
in subsequent sections we use scaled versions of $b_\ell,$ e.g., $b_\ell/16.$

If the functions of $\varepsilon$ in the asymptotic expansion~(\ref{asym}) are not known,
a nonlinear extrapolation or convergence acceleration may be
suitable~\cite{wynn56,sidi79b,levin81,ford87}.
As an example of a nonlinear extrapolation method,
the $\epsilon$-algorithm~\cite{wynn56} implements the sequence-to-sequence transformation
by~\cite{shanks55} recursively;
and can be applied when the $\varphi$ functions are of the form
$\varphi_k(\varepsilon) = \varepsilon^{\beta_k} \log^{\nu_k}(\varepsilon),$
under some conditions on $\nu_k$ and $\beta_k$
and if a geometric sequence is used for $\varepsilon.$
The actual form of the underlying $\varepsilon$-dependency does not need to be specified.

Table~\ref{epsalg} gives the recurrence of the $\epsilon$ algorithm 
for a sequence $S_\kappa, ~\kappa = 0,1,\ldots$ and
depicts the layout of the computations in a triangular table. 
\begin{table}
\begin{center}
\begin{minipage}{3in}
\begin{tabular}{ccccc}
& $\tau_{00}$ & & & \\
0 & & $\tau_{01}$ & & \\
& $\tau_{10}$ & & $\tau_{02}$ & \\
0 & & $\tau_{11}$ & &\ldots  \\
& & \ldots & & \ldots\\
& & \ldots & & \ldots\\
0 & & $\tau_{\kappa-1,1}$ & & \ldots \\
& $\tau_{\kappa 0}$ & & $\tau_{\kappa-1,2}$ & \\
0 & & $\tau_{\kappa 1}$ & \\
& $\tau_{\kappa+1,0}$ & & & \\
\end{tabular}
\end{minipage}
\begin{minipage}{2in}
\begin{align}
\tau_{\kappa,-1} &= 0 \nonumber \\
\tau_{\kappa, 0} &= S_\kappa \nonumber\\
\tau_{\kappa,\lambda+1} &= \tau_{\kappa+1,\lambda+1} + \frac{1}{\tau_{\kappa+1,\lambda}-\tau_{\kappa \lambda}}\nonumber
\end{align}
\end{minipage}
\end{center}
\caption{$\epsilon$-algorithm table}
\label{epsalg}
\end{table}

\section{Hypergeometric function}
\label{hypergeometric}
A representation of the hypergeometric function is given by the Gauss series
$$
F(a,b,c;z) = \sum_{k=0}^\infty \frac{\Gamma(a+k) \,\,\Gamma(b+k)}{\Gamma(c+k)}\,                               \frac{z^k}{k!}
$$
which has $|\,z\,| = 1$ as its circle of convergence
and has an analytic continuation defined by the Euler integral~\cite{abramowitz65},
\begin{equation}
{}_2F_1\,(a,b,c;z) = \frac{\Gamma(c)}{\Gamma(b) \,\Gamma(c-b)}
             \int_0^1 ~\frac{t^{\,{b-1}}(1-t)^{\,{c-b-1}}}{(1-t\,z)^{\,a}} ~dt,
\label{int2f1}
\end{equation}
$Re \,c > \,Re \,b > 0,$ which denotes a one-valued analytic 
function in the complex plane cut along the real axis from 1 to $\infty.$


For a numerical computation where $z \in \mathbb{R},$ we replace $z$ by $z+i\delta$ 
and evaluate the limit of ${}_2F_1\,(a,b,c;z+i\delta)$ as $\delta \rightarrow 0$
by solving linear systems of the form~(\ref{exp}) with $\varphi_k(\delta_\ell) = \delta_\ell^k,
 ~k = 1,\ldots,\nu$ and \,$S(\delta_\ell) = {}_2F_1\,(a,b,c;z+i\delta_\ell), 
 ~\ell = 0,\ldots,\nu.$

Table~2 lists results for a problem
set from~\cite{kurihara06} where $a = l+1, ~b = l+m, ~c = l+m+n+1$ 
at $z = 10$ (real). The relative error tolerances for the outer and inner integration
are set at $10^{-25}$ and $10^{-26},$ respectively. Since $b$ and $c$ are positive 
integer, the numerator in the integrand is polynomial, so there are no end-point 
singularities. We use the general adaptive integrator {\sc Dqag} of {\sc Quadpack}
for the iterated integrations, with the 7-point Gauss and 15-point Kronrod
rules applied on each subinterval in its adaptive partitioning strategy.

Note that the weights and abscissae of the integration rules in {\sc Dqag} are
given to 33 digits and the relative machine accuracy in quadruple precision is
of about the same order. Table~2 gives quadruple precision results, 
extending the (10-digit) accuracy of the double precision calculations reported in~\cite{jocs11}.
As compiler we use the intel Fortran Composer XE with the -r16 flag. The calculations
are run on a Macbook-Pro laptop with 3.06 GHz Intel Core 2 Duo processor and 8 GB memory
(Mac OS X Version 10.6.4).

\begin{table}
\centering{
\begin{scriptsize}
\begin{tabular}{|c|c|c|c|c|c|c|c|}\hline
  \multicolumn{3}{|c|}{\sc Parameters}
  & {\sc Real/} & {\sc Extrapolated} & {\sc Estim.\,Rel.} & & {\sc Total Int.} \\
\raisebox{0.05in}{l} & \raisebox{0.05in}{m} & \raisebox{0.05in}{n} & \raisebox{0.05in}{\sc Imag.} & \raisebox{0.05in}{{\sc Result} ${\mathcal S}^{(\nu)}$} & \raisebox{0.05in}{\sc Err. $E_r^{(\nu)}$} & \raisebox{0.1in}{$\nu$} & \raisebox{0.05in}{\sc Time (s)}\\
\hline
 1 & 1 & 1 & {\sc Real}  & -1.4533220287861469626056457257e-02 & 7.09e-29 & 22 & 7.58e-02 \\
   &   &   & {\sc Imag.} & -1.5079644737231007544620688243e-01 & 2.34e-28 & 22 & 9.70e-02 \\
 1 & 2 & 3 & {\sc Real}  &  8.4177671687810393005426755543e-02 & 2.85e-28 & 22 & 1.26e-01 \\
   &   &   & {\sc Imag.} & -2.2902210444669592708392670265e-01 & 1.61e-29 & 22 & 1.20e-01 \\
 2 & 1 & 1 & {\sc Real}  &  1.08766468837090905178932142e-02   & 7.26e-26 & 21 & 1.08e-01 \\
   &   &   & {\sc Imag.} &  2.63893782901542632030862046e-02   & 1.66e-27 & 22 & 1.30e-01 \\
 2 & 3 & 4 & {\sc Real}  & -2.89056808231506117361219566e-02   & 2.38e-26 & 22 & 1.47e-01 \\
   &   &   & {\sc Imag.} &  5.57897846432151278376445447e-02   & 1.44e-27 & 22 & 1.42e-01 \\
 3 & 1 & 2 & {\sc Real}  & -1.02672179830170273331993276e-02   & 2.94e-27 & 21 & 2.42e-01 \\
   &   &   & {\sc Imag.} & -5.65486677646162782923279e-02      & 4.29e-24 & 21 & 2.70e-01 \\
 3 & 4 & 5 & {\sc Real}  &  8.12135882438810010895282e-03      & 2.58e-24 & 21 & 1.79e-01 \\
   &   &   & {\sc Imag.} & -1.27463626362435298394122e-02      & 3.92e-25 & 23 & 2.13e-01 \\
\hline
\end{tabular}
\end{scriptsize}
\caption{Integration and extrapolation results for ${}_2F_1(l+1,l+m,l+m+n+1,z+i0)$
for relative integration error tolerance of $10^{-25}$ and the
Bulirsch sequence for extrapolation}
}
\label{hyper}
\end{table}
According to~(\ref{exp}) we obtain ${\mathcal S}^{(\nu)} \approx a_0$ for the systems of
order $(\nu+1)\times (\nu+1), ~\nu = 1,2,\ldots$. The difference of successive results
in \,$E_a^{(\nu)} = |{\mathcal S}^{(\nu)}-{\mathcal S}^{(\nu-1)}|$ gives a measure of convergence.
Table~2 lists $l,\, m,\, n$ followed by the result ${\mathcal S}^{(\nu)},$
the relative measure of convergence $E_r^{(\nu)} = E_a^{(\nu)}/\,|{\mathcal S^{(\nu)}}|,$ 
the value of $\nu$ and the time taken (in seconds) for the integrals $S(\delta_\ell),
 ~\ell=0,\ldots,\nu,$ needed to obtain the result.
The other times in the program are not taken into account, including the system
solving time which was measured and found negligible compared to the integration times.

While we could have listed the result obtained after the error falls below $10^{-25},$
we instead listed the result with the smallest estimated relative error.
As expected, better accuracy is obtained for the smaller values of \,$l, \,m, \,n.$

\section{Asymptotics for one off-shell, two on-shell particles}
\label{one-off-shell}
We first address the IR divergent integral $J_3\,(p_1^2,p_2^2,p_3^2;n_x,n_y)$ from~\cite{kurihara05a}
with one off-shell ($p_3^2 \ne 0$) and two on-shell particles ($p_1^2 = p_2^2 = 0$),
which we denote here by $J_3\,(p_1^2,p_2^2,p_3^2;n_x,n_y;\varepsilon) = \frac{1}{(4\pi)^2} \,I_{3}^{n_x,n_y}(\varepsilon),$
 ~with
\begin{align}
I_{3}^{n_x,n_y}(\varepsilon) &= \frac{\varepsilon \,\Gamma(-\varepsilon)}
   {(4\pi \mu_R^2)^{\,\varepsilon}} \int_0^1 dx \int_0^{1-x} dy\,
   \frac{x^{n_x}y^{n_y}}{(-p_3^2 xy-i0)^{\,1-\varepsilon}} \label{Ij3} \\
   &= \varepsilon \,\Gamma(-\varepsilon) \,
   \left( \frac{-\tilde{p_3}^2}{4\pi \mu_R^2} \right)^{\,\varepsilon}\frac{1}{-p_3^2}\,
   \frac{B(n_x+\varepsilon,n_y+\varepsilon)}{n_x+n_y+2\varepsilon}. \nonumber
\end{align}
where $\tilde{p_3}^2 = p_3^2+i0$ \,and
$\mu_R^2$ is a renormalization constant (which we will replace by
$\mu_R^2 \leftarrow e^{\gamma_E}/(4\pi),$ $\gamma_E$ is Euler's constant).
The introduction of the parameter $\varepsilon$ pertains to the 
dimension regularization technique~\cite{muta08}.

When $n_x = \eta \ne 0$ or $n_y \ne 0,$ we have
\begin{equation}
I_{3}^{\,\eta,0}(\varepsilon) \sim  \frac{1}{p_3^2}\,\left(\frac{C_{-1}}{\varepsilon} + C_0 + {\mathcal O}(\varepsilon)\,\right) ~~~~\mbox{with}
\label{Lvertexnx}
\end{equation}
$$C_{-1} = \frac{1}{\eta}, ~~~~C_0 = -\frac{2}{\eta^2}+\frac{1}{\eta}\,\left(\ln(-p_3^2) -\sum_{j=1}^{\eta-1}\frac{1}{j}\,\right).$$
A linear extrapolation can be formulated using systems of the form~(\ref{exp}) with
$S_\ell(\varepsilon_\ell) = \varepsilon_\ell \hat{I}(\varepsilon_\ell)$ where
$\hat{I}(\varepsilon_\ell) \approx I_{3}^{\,\eta,0}(\varepsilon)$ and $\varphi_k(\varepsilon_\ell) =
\varepsilon_\ell^k.$
\begin{table}\centering
\begin{scriptsize}
\begin{tabular}{|c|r|c|c|c|}\hline
 & \multicolumn{2}{c|}{{\sc Integration}}
& \multicolumn{2}{c|}{{\sc Extrapolation Results}} \\ \raisebox{1mm}{$\nu$} & {\sc \#~Evals} & {\sc Time} & $C_{-1}$ & $C_0$ \\
& & \raisebox{1mm}{(s)} & & \\
\hline
1 & 1125075 & 1.10 & 0.8707294126792779500928011356924 & -0.5511174754172542062710316596 \\
2 & 1349535 & 1.34 & 0.3457684478555631741492957488633 &  3.1236092783487492253335060482 \\
3 & 1348875 & 1.33 & 0.5262024637357823738306365020734 &  0.7779670719058996294760762565 \\4 & 134
9355 & 1.33 & 0.4976817315764365660976698223580 &  1.3769024472521615918683765304 \\
5 & 1324125 & 1.30 & 0.5001453140694736901417736134445 &  1.3030216329875385877494858112 \\
6 & 1274625 & 1.26 & 0.5000001921131322347567104032762 &  1.3025651116288535549087530061 \\7 & 127
4625 & 1.25 & 0.4999999959467322713652940448364 &  1.3025850791321435704903401609 \\
8 & 1225125 & 1.21 & 0.4999999999993260163866288358759 &  1.3025850932043285642085160900 \\
9 & 1200375 & 1.18 & 0.5000000000000053069685193174512 & 1.3025850929940632251166716065 \\10 & 1175625 & 1.16 & 0.4999999999999999722450728860463 & 1.3025850929940632251166716065 \\
11 & 1101375 & 1.08 & 0.5000000000000000001089023336541 & 1.3025850929940455873126312708 \\
12 & 1076625 & 1.06 & 0.4999999999999999999997158886319 & 1.3025850929940456843793808956 \\
13 & 1037475 & 1.02 & 0.5000000000000000000000005564493 & 1.3025850929940456840169987640 \\
14 & 1037475 & 1.00 & 0.4999999999999999999999999992988 & 1.3025850929940456840179932776 \\
15 & 952425 & 0.90 & 0.4999999999999999999999999999980 & 1.3025850929940456840179914927 \\
\hline
\multicolumn{2}{|r|}{\em Total time :} & 20.31 & \hspace*{-1.4cm}{\em Exact :~~~~~~~~~~~~} 0.5 & 1.3025850929940456840179914547 \\
\hline
\end{tabular}
\end{scriptsize}\caption{Integration performance ({\sc Dqags}$)^2,$
for rel. integration error
tolerances $5\times 10^{-24}$ (outer), $10^{-25}$ (inner); $n_x = \eta = 2, ~n_y = 0$
and $p_3^2 = 100.$ Extrapolated real values.}
\label{table2}
\end{table}

Table~\ref{table2} displays results for the real parts of the coefficients $C_{-1}$ and 
$C_0$ in the expansion, for $n_x = \eta = 2$ and $n_y = 0.$ 
In view of the integrand singularity along $y = 0$ through the factor $y^{\varepsilon-1}$
where the exponent approaches -1, and a second derivative singularity along $x = 0$ 
through the factor $x^{1+\varepsilon}$ in~(\ref{Ij3}), we perform the
calculation of $\hat{I}(\varepsilon_\ell)$ with the program {\sc Dqags} of {\sc Quadpack}~\cite{pi83}, which is
equipped to deal with these end-point singularities. {\sc Dqags} uses the 7-point Gauss and 15-point Kronrod
rule on each subinterval created in its adaptive subdivision strategy.

We set the requested relative accuracies to $5\times 10^{-24}$ for the outer and
$10^{-25}$ for the inner integral. Table~\ref{table2} lists the sequence of extrapolation 
results and shows how much accuracy can be reached.
(After that, the accuracy no longer improves; i.e., stagnates for a few steps and 
decreases). 
The difference between successive results provides a good estimate of the convergence. 
For this problem it usually gives a somewhat conservative bound for the actual error,
in the sense that the difference with the result of the previous step is in fact a measure
of the error in the previous step. 
The final accuracy reached for $C_{-1}$ is about $2\times 10^{-30}$ (absolute error), 
$4 \times 10^{-30}$ (relative error). The accuacy in $C_0$ lags behind and reaches
about $4\times 10^{-26}$ (absolute) or $3\times 10^{-26}$ (relative error). 

When $n_x = n_y = 0$ we have
\begin{equation}
I_{3}^{\,0,0}(\varepsilon) \sim \frac{1}{p_3^2} \,\left( \frac{C_{-2}}{\varepsilon^2}
                                  +\frac{C_{-1}}{\varepsilon}+C_0
                                  +{\mathcal O}(\varepsilon)\,\right) ~~~~\mbox{with} \nonumber
\end{equation}
\begin{equation}
C_{-2} = 1, ~~~C_{-1} = \ln(-p_3^2), ~~~C_0 = -\frac{\pi^2}{12}+\frac{1}{2}\ln^2(-p_3^2).
\label{Lvertex00}
\end{equation}
With relative tolerances of $10^{-26}$ for the outer and $5\times 10^{-27}$ 
for the inner integrations, we obtain the best relative accuracies for $C_{-2}, ~C_{-1}$ and
$C_0$ at $\nu = 15,$ of 7.88e-25, ~4.33e-22 and 4.0e-19, respectively.
Note that the best accuracies for this problem obtained in double precision with integration tolerances
of $10^{-13}$ (outer), $5\times 10^{-14}$ (inner) are reported in~\cite{jocs11} as
3.06e-12, 1.44e-10 and 7.99e-09 at $\nu = 11.$ 
The latter can also be compared with our $10^{-26},$ $5\times 10^{-27}$ quadruple precision
result at $\nu = 11,$ which yields 3.18e-14, ~3.04e-12 and 5.24e-10, respectively.

\section{Asymptotics for one on-shell, two off-shell particles}
\label{two-off-shell}
In the case of one on-shell ($p_1^2 = 0$) and two off-shell ($p_2^2 \ne 0, ~p_3^2 \ne 0$) particles,
the IR divergent integral is
\begin{align}
J_3\,(0,p_2^2,p_3^2;n_x,n_y;\varepsilon) 
 &= \frac{1}{(4\pi)^2} \,
   \, J_3\,(0,0,p_3^2;n_x,n_y;\varepsilon) ~ \,{}_2F_1\,(1,1-\varepsilon,2+n_x;\frac{p_3^2-p_2^2}{\tilde{p_3}^2}) ~\,\frac{n_x+\varepsilon}{n_x+1} \label{pr} \\
 &= \frac{1}{(4\pi)^2} \,
   \frac{\varepsilon \,\Gamma(-\varepsilon)}
   {(4\pi \mu_R^2)^{\,\varepsilon}} \int_0^1 dx \int_0^{1-x} dy\,
   \frac{x^{n_x}y^{n_y}}{(-(p_3^2-p_2^2)xy-p_2^2 y\,(1-y)-i0)^{\,1-\varepsilon}
}. \label{Ij23} 
\end{align}
For $n_x = n_y = 0$ we use the expansion
\begin{equation}
J_3\,(0,p_2^2,p_3^2;n_x,n_y;\varepsilon) \sim \frac{\tilde{C}_{-1}}{\varepsilon} 
  + \tilde{C}_0 + {\mathcal O}(\varepsilon)
\label{v23}
\end{equation}
with 
\begin{align}
 &\tilde{C}_{-1} = -\frac{1}{(4\pi)^2 p_3^2} \,\frac{\ln(1-z)}{z} \nonumber \\
 &\tilde{C}_0 = -\frac{1}{(4\pi)^2 p_3^2} \,\left(\,C_{-1} \frac{\ln(1-z)}{z} + 
                                         \,\frac{\ln^2(1-z)}{2z}\,\right) \nonumber
\end{align}
with \,$z = \frac{p_3^2-p_2^2}{\tilde{p_3}^2}$\, and the $C_{-1}$ coefficient of~(\ref{Lvertex00}).
In the subsections below we will give results for two
methods 
for the computation of $J_3\,(0,p_2^2,p_3^2;n_x,n_y;\varepsilon),$ the
first based on~(\ref{pr}) and the second on~(\ref{Ij23}).

\subsection{Computation of $J_3\,(0,p_2^2,p_3^2;n_x,n_y)$ with hypergeometric function}
\label{two-off-shell-hypergeometric}
We intend to perform an extrapolation
with respect to the parameter $\varepsilon$ for dimensional regularization.
According to~(\ref{pr}), each term in the extrapolation sequence consists of a double integral multiplied with a hypergeometric
function, both dependent on $\varepsilon.$ We calculate the hypergeometric function
numerically using the $\delta$ extrapolation of Section~\ref{hypergeometric}, to
alleviate the characteristic non-integrable singularity on the real axis when it
occurs inside the integration interval.

Here $z = \frac{p_3^2-p_2^2}{\tilde{p_3}^2}$ in the fourth argument of ${}_2F_1$ 
in~(\ref{pr}) is replaced by $z+i\delta.$ Thus in this process we need to evaluate 
sequences of hypergeometric functions of the form
$$\,{}_2F_1\,\,(1,1-\varepsilon_\ell,\,2+n_x;\,\frac{p_3^2-p_2^2}{\tilde{p_3}^2}+i\,\delta_\kappa\,), ~~\kappa = 1,2,\ldots$$
for $\varepsilon = 1,2,\ldots.$
The number of $\kappa$-values needed depends on the convergence for each (fixed) $\varepsilon.$
As the exponents \,$b-1 = -\varepsilon_\ell$\, and \,$c-b-1 = n_x+\varepsilon_\ell$\, in 
the integrand numerator $t^{b-1}(1-t)^{c-b-1}$ of the Euler integral~(\ref{int2f1})
are non-integer, we use the {\sc Quadpack} program {\sc Dqags} to treat the integrand
behavior at the end-points.

Extrapolated (real part) results for $\tilde{C}_{-1}$ and $\tilde{C}_0$ in~(\ref{v23})
and actual absolute errors are listed in Table~\ref{table4}. 
By setting $p_2^2 = 40, ~p_3^2 = -100,$ for the purpose of a numerical example,
we have that $Re(z) = 1.4,$ so that the hypergeometric function has an integrand singularity
at $t = 1/1.4$ in the interior of the integration interval $(0,1).$
\begin{table}
\centering
\begin{scriptsize}
\begin{tabular}{|c|c|c|c|c|c|}\hline
 & {\sc Time} & \multicolumn{4}{c|}{{\sc Extrapolation Results}} \\
\raisebox{1mm}{$\nu$} & {\sc Int.(s)} & $\tilde{C}_{-1}$ & {\sc |Error|} & $\tilde{C}_0$ & {\sc |Error|} \\
\hline
4  & 3.47 & -4.2192812666950192419664334e-05 & 7.47e-07 & -3.7121279333017303070237e-04 & 2.39e-05 \\
5  & 3.47 & -4.1395889404540049511801909e-05 & 5.04e-08 & -3.9751126098970774666184e-04 & 2.42e-06 \\
6  & 3.41 & -4.1448485840565542759037853e-05 & 2.21e-09 & -3.9493403562445857754728e-04 & 1.59e-07 \\
7  & 3.39 & -4.1446205836959577424975383e-05 & 7.16e-11 & -3.9510047588769404693383e-04 & 7.45e-09 \\
8  & 3.33 & -4.1446278988182334267440660e-05 & 1.53e-12 & -3.9509279500930457847498e-04 & 2.32e-10 \\
9  & 3.30 & -4.1446277437326639980970560e-05 & 2.43e-14 & -3.9509303229022580430491e-04 & 5.25e-12 \\
10 & 3.53 & -4.1446277461859729880198253e-05 & 2.56e-16 & -3.9509302696654529617250e-04 & 7.97e-14 \\
11 & 3.86 & -4.1446277461602157717453434e-05 & 2.01e-18 & -3.9509302704716538311163e-04 & 8.86e-16 \\
12 & 3.75 & -4.1446277461604182432081146e-05 & 1.05e-20 & -3.9509302704627248396080e-04 & 6.66e-18 \\
13 & 3.81 & -4.1446277461604171849957952e-05 & 4.14e-23 & -3.9509302704627918244479e-04 & 3.67e-20 \\
14 & 3.77 & -4.1446277461604171891430927e-05 & 1.08e-25 & -3.9509302704627914557531e-04 & 1.37e-22 \\
15 & 3.81 & -4.1446277461604171891322514e-05 & 3.19e-28 & -3.9509302704627914571332e-04 & 5.44e-25 \\
16 & 3.76 & -4.1446277461604171891322881e-05 & 4.87e-29 & -3.9509302704627914571267e-04 & 1.12e-25 \\
17 & 3.82 & -4.1446277461604171891322692e-05 & 1.39e-28 & -3.9509302704627914571315e-04 & 3.69e-25 \\
18 & 3.78 & -4.1446277461604171891323096e-05 & 2.63e-28 & -3.9509302704627914571171e-04 & 1.07e-24 \\
19 & 3.82 & -4.1446277461604171891322859e-05 & 2.63e-29 & -3.9509302704627914571292e-04 & 1.40e-25 \\
\hline
\multicolumn{2}{|c|}{~~~~~~~~~~~\emph{Ex :}} & -4.1446277461604171891322832e-05 &  & -3.9509302704627914571278e-04
 & \\
\hline\end{tabular}
\end{scriptsize}
\caption{Integration performance ({\sc Dqags}$)^2,$
for rel. integration error tolerances $10^{-26}$ (outer), $5\times 10^{-27}$ (inner);
vertex with one on-shell, two off-shell particles,
$n_x = n_y = 0$ and $p_2^2 = 40, ~p_3^2 = -100.$
Extrapolated real values and corresponding actual absolute errors.}
\label{table4}
\end{table}

The relative error tolerances
for {\sc Dqags} were $10^{-26}$ for the outer integration and $5\times 10^{-27}$ for the
inner integrals. The maximum number of subdivisions was set to 150 for $J_3$ in both 
directions and to 300 for ${}_2F_1.$ 
For the dimensional regularization extrapolation, the Bulirsch sequence was used
starting at 3, i.e., 3,4,6,\ldots.

For $p_3^2 = -100,$ the integral in~(\ref{Ij3}) is real and is multiplied with the
(complex) value of the hypergeometric function. The real and imaginary parts of the
latter are currently computed separately which, for the real part, took between about
0.44 and 0.55 seconds, and for the imaginary part between 0.33 and 0.41 seconds. Thus the
time for the hypergeometric function computation for each equation in the linear system 
is under a second and is dominated by the times for the integrations of 
$J_3$ in~(\ref{Ij3}) (given in Table~\ref{table4}). The total time spent in the double
integration is 68.3 seconds (including that of the first three steps, which are not shown).

\begin{figure}[t]
\vspace*{-0.4in}
\centering
\includegraphics[clip,width=0.5\linewidth]{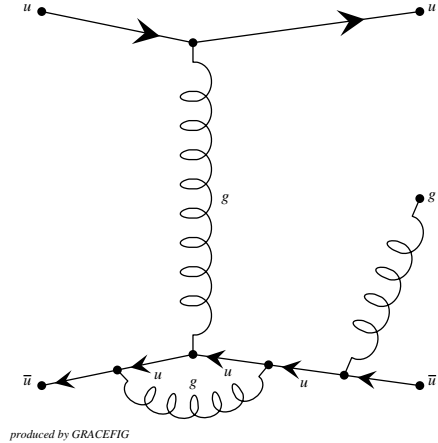}
\caption{$qq~ \rightarrow ~qqg$ diagram. The vertical gluon propagator carries $p_3^2$ and
the u-quark emitting a gluon carries $p_2^2.$
}
\label{qqqqg}
\end{figure}

A sample diagram for the vertex correction with one on-shell and two
off-shell particles is depicted in Figure~\ref{qqqqg} for the in $qq~ \rightarrow ~qqg$  interaction.
Here the vertical gluon propagator carries $p_3^2$ and
the u-quark emitting a gluon carries $p_2^2.$ Note that the vertical gluon
propagator can be virtual so that the square of its momentum becomes negative.

\subsection{Computation of $J_3\,(0,p_2^2,p_3^2;n_x,n_y)$ using direct integration} 
\label{two-off-shell-direct}
\begin{table}[b]
\centering
\begin{scriptsize}
\begin{tabular}{|c|c|c|c|c|}\hline
$\nu$ & $C_{-1}$ & {\sc |Error|} & $C_0$ & {\sc |Error|} \\
\hline
 4 & -4.2192812666986e-05 & 7.47e-07 & -3.712127933292e-04 & 2.39e-05 \\
 5 & -4.1395889404493e-05 & 5.04e-08 & -3.975112609915e-04 & 2.42e-06 \\
 6 & -4.1448485840595e-05 & 2.21e-09 & -3.949340356225e-04 & 1.59e-07 \\
 7 & -4.1446205836943e-05 & 7.16e-11 & -3.951004758891e-04 & 7.45e-09 \\
 8 & -4.1446278988175e-05 & 1.53e-12 & -3.950927950097e-04 & 2.23e-10 \\
 9 & -4.1446277437321e-05 & 2.43e-14 & -3.950930322905e-04 & 5.24e-12 \\
10 & -4.1446277461827e-05 & 2.23e-16 & -3.950930269725e-04 & 7.38e-14 \\
11 & -4.1446277461600e-05 & 3.96e-18 & -3.950930270436e-04 & 2.69e-16 \\
\hline
\emph{Ex:} & -4.1446277461604e-05 &  & -3.950930270463e-04 & \\
\hline
\end{tabular}
\end{scriptsize}
\caption{Integration performance ({\sc Dqags}$)^2,$
for rel. integration error tolerances $10^{-15}$ (outer), $10^{-16}$ (inner); 
Vertex with one on-shell, two off-shell particles,
$n_x = n_y = 0$ and $p_2^2 = 40, ~p_3^2 = -100.$ Extrapolated real parts 
and actual absolute errors.}
\label{table5}
\end{table}
In this section we calculate the integral in~(\ref{Ij23}) directly, which becomes highly
singular at $y = 0$ as $\varepsilon\rightarrow 0$ and non-integrable for $\varepsilon = 0,$
in view of the factor 
$y^{\varepsilon-1}$ in the integrand $\frac{y^{\varepsilon-1}}{(-D)^{1-\varepsilon}}.$ 
It can be seen that $(p_3^2-p_2^2)x+p_2^2(1-y) = 0$ along a line that goes through
$(\,p_2^2\,/\,(p_2^2-p_3^2)\,,0)$ and $(0,1).$ Thus for $p_2^2 = 40, ~p_3^2 = -100,$ 
this line runs through the integration domain, where it causes a singularity that
becomes non-integrable as $\varepsilon \rightarrow 0.$ 

\begin{table}[b]
\centering
\begin{scriptsize}\begin{tabular}{|c|c|c|c|c|}\hline
$\nu$ & $C_{-1}$ & {\sc |Error|} & $C_0$ & {\sc |Error|} \\
\hline
 1 &  ~0.758010428109e-04\, & \,6.63e-05 &  ~9.62356140022e-04\, & \,4.38e-04 \\
 2 &  ~1.464854199104e-04\, & \,4.38e-06 &  ~4.67565500325e-04\, & \,5.66e-05 \\
 3 &  ~1.420636477897e-04\, & \,3.90e-08 &  ~5.25048537895e-04\, & \,4.91e-07 \\
 4 &  ~1.421000460091e-04\, & \,2.58e-09 &  ~5.24284175287e-04\, & \,8.47e-08 \\
 5 &  ~1.421026578481e-04\, & \,3.01e-11 &  ~5.24197984601e-04\, & \,1.48e-09 \\
 6 &  ~1.421026279735e-04\, & \,2.13e-13 &  ~5.24199448458e-04\, & \,1.55e-11 \\
 7 &  ~1.421026277585e-04\, & \,2.16e-15 &  ~5.24199464154e-04\, & \,2.32e-13 \\
 8 &  ~1.421026277612e-04\, & \,5.84e-16 &  ~5.24199463865e-04\, & \,5.64e-14 \\
\hline
\emph{Ex:} &  1.421026277606e-05 &  &  5.24199463922e-04 & \\
\hline
\end{tabular}
\end{scriptsize}
\caption{Integration performance ({\sc Dqags}$)^2,$
for rel. integration error tolerances $10^{-15}$ (outer), $10^{-16}$ (inner);
Vertex with one on-shell, two off-shell particles,
$n_x = n_y = 0$ and $p_2^2 = 40, ~p_3^2 = -100.$ Extrapolated imaginary parts
and actual absolute errors.}
\label{table6}
\end{table}
According with previous work (e.g., ~\cite{edcpp03,eddacat03,loopfestV,acat08,ddiccsa10}), 
we will replace $i0$ by $i\delta$ in the integrand denominator of~(\ref{Ij23}). 
This leads to integrals of the form
\begin{align}
I_3\,(\varepsilon,\delta) = & \frac{1}{(4\pi)^2} \,
   \frac{\varepsilon \,\Gamma(-\varepsilon)}
   {(4\pi \mu_R^2)^{\,\varepsilon}} \int_0^1 dx \int_0^{1-x} dy\,
   \frac{x^{n_x}y^{n_y}}{(-(p_3^2-p_2^2)xy-p_2^2 y\,(1-y)-i\delta)^{\,1-\varepsilon}
} \nonumber \\
    & \sim \frac{\tilde{\tilde{C}}_{-1}(\delta)}{\varepsilon} + \tilde{\tilde{C}}_0(\delta)  + {\mathcal O}(\varepsilon)
\label{I3}
\end{align}
to be computed in an extrapolation as $\delta\rightarrow 0,$ for each 
(fixed) value of the dimensional regularization parameter $\varepsilon.$
The expansion~(\ref{v23}) or~(\ref{I3}) with respect to $\varepsilon$ justifies a linear extrapolation.
However, a linear extrapolation as $\delta\rightarrow 0$ is not suited, as
it cannot be assumed that the underlying expansion is in integer powers of $\delta.$

Preliminary results are shown in Tables~\ref{table5}-\ref{table6} for the real and the imaginary
parts, respectively. 
We employ a sequence of $\delta_\kappa = 2^{-8-\kappa}, ~\kappa = 0,1,\ldots$ in an extrapolation 
using the $\epsilon$-algorithm of Wynn~\cite{shanks55,wynn56}.
The current implementation incorporates a simple strategy with an extrapolation sequence of 
fixed length (= 18), which should be improved with a termination criterion based on the estimated errors.
Geometric sequences of $\delta$ with a smaller ratio should also be tested.
For the extrapolation with respect to $\varepsilon$ the Bulirsch sequence $3,4,6,\ldots$ is used.

The requested accuracies for {\sc Dqags} are $10^{-15}$ for the outer and $10^{-16}$ for the inner 
integrations.
The real and imaginary parts are calculated separately as the {\sc Quadpack} programs
currently do not handle complex functions. 
The imaginary parts converge more quickly than the real parts for this problem. 
We list the steps 1 to 8; the last for a $9\times 9$ system. 
The results do not improve after that.
Note that it is not necessary to solve the previous systems,
but the integrals need to be calculated to set up the system.
Successive systems can be solved to provide an estimate of the error.


\section*{Conclusions}
\noindent
We use 1D integration programs from {\sc Quadpack} for the iterated integrations underlying
the extrapolation processes for IR divergent vertex integrals. 
We give preliminary results in quadruple precision which demonstrate that the calculations 
result in high accuracy.
For the vertex computation with one on-shell and two off-shell particles we devise a double 
extrapolation, since problems are introduced with integrand singularities in the interior
of the domain as well as on the boundaries.
Many integrals result, which makes the computation a good candidate 
for parallelization, especially when applied to more complicated diagrams and possibly
in higher precision. 

More work is needed for improvements and extensions of these strategies and analysis of the 
numerical methods and results.
This computation is a step toward a more automatic numerical handling of various types of 
loop integrals, thereby circumventing the need for a precise knowledge of the location
and structure of the singularities.




\end{document}